\documentclass[%
 reprint,
superscriptaddress,
 amsmath,amssymb,
 aps,
]{revtex4-2}

\usepackage{graphicx}
\usepackage{dcolumn}
\usepackage{bm}

\usepackage{soul}
\usepackage{todonotes}
\usepackage{comment}

\begin{document}

\preprint{APS/123-QED}

\title{Transition time of  a bouncing drop}

\author{Yahua Liu}
\thanks{These authors contributed equally to this work}
\affiliation{State Key Laboratory of High-Performance Precision Manufacturing, Dalian University of Technology, Dalian 116024, China.}

\author{Seyed Ali Hosseini}
\thanks{These authors contributed equally to this work}
\affiliation{Department of Mechanical and Process Engineering, ETH Zurich, 8092 Zurich, Switzerland.}

\author{Cong Liu}
\affiliation{State Key Laboratory of High-Performance Precision Manufacturing, Dalian University of Technology, Dalian 116024, China.}

\author{Milo Feinberg}
\affiliation{Department of Mechanical and Process Engineering, ETH Zurich, 8092 Zurich, Switzerland.}

\author{Benedikt Dorschner}
\affiliation{Department of Mechanical and Process Engineering, ETH Zurich, 8092 Zurich, Switzerland.}

\author{Zuankai Wang}
\email{zk.wang@polyu.edu.hk}
\affiliation{Department of Mechanical Engineering, The Hong Kong Polytechnic University, Hong Kong 999077, China.}

\author{Ilya Karlin}
\email{ikarlin@ethz.ch}
\affiliation{Department of Mechanical and Process Engineering, ETH Zurich, 8092 Zurich, Switzerland.}

\date{\today}

\begin{abstract}
Contact time of bouncing drops is one of the most essential parameters to quantify the water-repellency of surfaces. Generally, the contact time on superhydrophobic surfaces is known to be Weber number-independent. Here, we probe an additional characteristic time, \emph{transition time} inherent in water drop impacting on superhydrophobic surfaces, marking a switch from a predominantly lateral to an axial motion. Systematic experiments and numerical simulations show that the transition time is also Weber number-independent and accounts for half the contact time.  Additionally we identify a Weber-independent partition of volume at the maximum spreading state between the rim and lamella and show that the latter contains 1/4 of the total volume of the drop.
\end{abstract}

\maketitle


\section{\label{sec:level1}Introduction}
Impact of drops on solid surfaces is a subject of interfacial hydrodynamics encountered in various applications, from spray-coating, ink-jet printing to cooling~\cite{josserand_drop_2016,yarin_drop_2006, chuanhua2017, wildeman_visser_sun_lohse_2016}. 
Of special interest is the strong repellence observed on superhydrophobic surfaces (SHSs), of particular importance for self-cleaning and anti-icing \cite{lv_bio-inspired_2014,kreder_design_2016, Stone2012, SemprebonButt2014, Kimliu2014, DengWang2020}.
The standard viewpoint represents drop bouncing off SHSs as a spread-recoil-rebound process. The contact time scales as the (square root of the) ratio between the mass of the droplet and surface tension, and is independent of the impact kinetic energy, or Weber number $\hbox{We}=\rho U_0^2 R_0 /\gamma$, where $U_0$, $R_0$, $\rho$ and $\gamma$ are impact velocity, radius, density and surface tension of the drop. An elegant explanation is based on a harmonic mode argument \cite{richard_contact_2002,okumura_water_2003}, in which the drop bouncing is mapped onto an equivalent spring-mass system. Recent studies also show that the contact time of bouncing droplets can be significantly reduced by controlling the surface macrotextures~\cite{bird_reducing_2013,gauthier_water_2015,wang_impact_2007,liu_pancake_2014,chantelot_water_2018}.

In the present work, we take a closer look at the dynamics of drop impacting SHSs through both experiments and simulations. We confirm previous observations on Weber-independence of contact time and further observe that the characteristic time marking a switch from a predominantly radial to an axial motion of the recoiling drop, termed transition time here, is also independent of Weber number and constitutes a half of the contact time on a flat SHS. This observation, based both identification thought the Worthington jet and the internal viscous dissipation rate of the drop, agrees very well with recent studies on the time evolution of the normal force exerted by drop onto the substrate~\cite{zhang2022impact}.
We further analyze the volume distribution at the maximum spreading state and observe the volume partition between the rim and lamella is also Weber independent.

The manuscript starts with an overview of both experimental and numerical techniques used in the context of this study. We then present measured transition times and present an in-depth analysis of the drop impact process through energy budget and flow field analyzes.

\section{\label{sec:methods}Methods}
\subsection{\label{sec:methods:experiments}Experiments}
\begin{figure}
	\centering
		\includegraphics[width=0.9\linewidth]{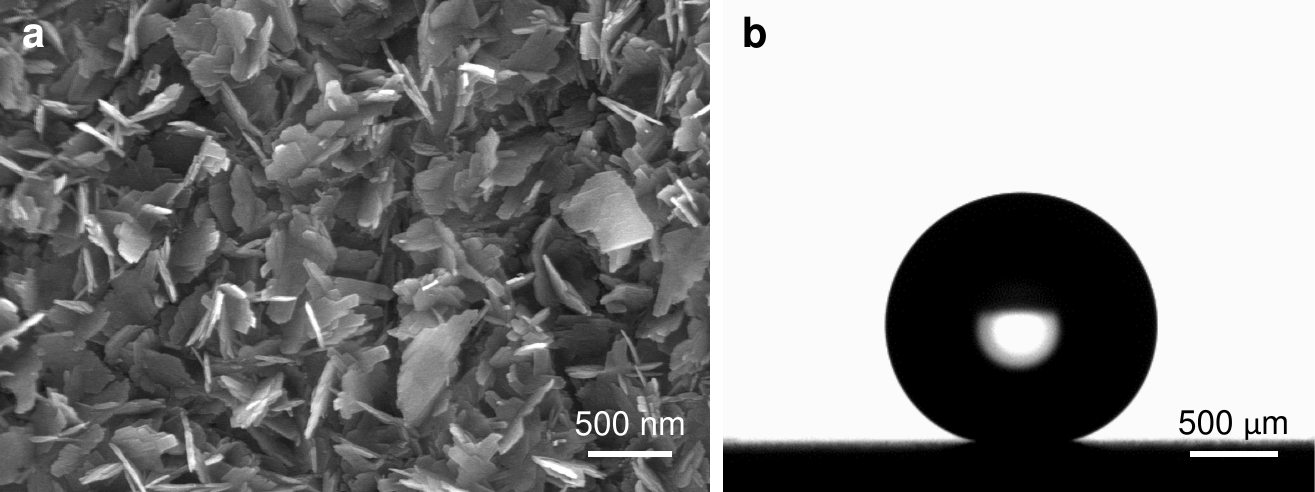}
	\caption{Surface characterization: (a) Scanning Electron Microscope images of the superhydrophobic surface. (b) Optical image showing a  $\sim 3 \mathrm{\mu L}$ water drop sitting on the surface.}
	\label{Figure1}
\end{figure}

Drop impact experiments were conducted on a superhydrophobic surface covered by nanosheets fabricated via chemical etching on a copper plate. Specifically, copper blocks with a size of $2\times2~{\rm cm}^2$ were first polished by fine sandpapers (1500\#), and then ultrasonically cleaned in ethanol and deionized water for 10~min, respectively, and dried in clean air, followed by immersion in an aqueous solution of 2.5~M/L sodium hydroxide and 0.1~M/L ammonium persulfate for 60~min. To render the surface superhydrophobic, the as-fabricated substrates were then put in 1 mM/L hexane solution of trichloro(1H,1H,2H,2H-perfluorooctyl)silane for $\sim$ 60 mins, followed by heat treatment at 150$^{\circ}$C in air for one hour. Scanning Electron Microscope image of the substrate is shown in Fig.~\ref{Figure1}. The process resulted in a surface contact angle $\theta=$ 166$^{\circ} \pm 3.2^{\circ}$.
\subsection{\label{sec:methods:lbm}Lattice Boltzmann solver for non-ideal fluids}
The simulations conducted in the present study were performed using the pseudo-potential lattice Boltzmann method (LBM)~\cite{shan_lattice_1993}. The discrete populations evolve in time and space and can be expressed as
{\small
\begin{flalign}
    &&
    f_i\left(\bm{r}+\bm{c}_i\delta t, t+\delta t\right) - f_i\left(\bm{r}, t\right) &= \frac{\delta t}{\tau}\left(f^{\rm eq}_i\left(\rho, \bm{u}\right) - f_i\left(\bm{r}, t\right)\right) & \nonumber \\
    && &\quad + f^{\rm eq}_i\left(\rho, \bm{u} + \frac{\bm{F}\delta t}{\rho}\right) & \nonumber \\
    && &\quad - f^{\rm eq}_i\left(\rho, \bm{u}\right), &
\end{flalign}}
\noindent where $\bm{c}_i$ ($i=1,\dots,Q$) and $\delta t$ are the discrete velocities and the discrete time-step size, respectively. We consider the standard $D3Q27$ lattice in three dimensions ($D=3$) with twenty-seven velocities ($Q=27$), $\bm{c}_i=(c_{ix},c_{iy},c_{iz})$, $c_{i\alpha}\in\{-1,0,1\}$. The relaxation time $\tau$ is tied to the local dynamic viscosity and density as $\tau=\mu/\rho \varsigma^2 + \delta t/2$ and the discrete equilibrium distribution function, in its full product form, is defined as~\cite{karlin_factorization_2010}

{\small
\begin{flalign}
    &&
    f^{\rm eq}_i\left(\rho, \bm{u}\right) =  \rho\prod_{\alpha=x,y,z} {\left(1-\mathcal{P}_{\alpha\alpha}\right)}^{1-\lvert c_{i,\alpha}\lvert} {\left(\frac{\mathcal{P}_{\alpha\alpha}+c_{i,\alpha}u_\alpha}{2}\right)}^{\lvert c_{i,\alpha}\lvert},
    &&
\end{flalign}}with
\begin{flalign}
    &&
    \mathcal{P}_{\alpha\alpha} = u_\alpha^2 + \varsigma^2,
    &&
\end{flalign}
where $u$ is the local fluid velocity and $\varsigma=\delta r/\sqrt{3}\delta t$ is the lattice sound speed. The body force $\bm{F}$ is defined as~\cite{sbragaglia_generalized_2007}
\begin{flalign}
    &&
    \bm{F} = -\psi(\bm{r})\sum_{i=1}^{Q} \frac{w_i}{\varsigma^2 \delta t} \bm{c}_i \left[ \mathcal{G}_1 \psi\left(\bm{r}+\bm{c}_i\delta t\right) + \mathcal{G}_2 \psi\left(\bm{r}+2\bm{c}_i \delta t\right)\right],
    &&
\end{flalign}
where the potential $\psi$ is computed as
\begin{flalign}
    &&
    \psi = \sqrt{\frac{2\left(P-\rho\varsigma^2\right)}{\mathcal{G}_1+2\mathcal{G}_2}}.
    &&
\end{flalign}

The parameters $\mathcal{G}_1$ and $\mathcal{G}_2$ in Eq. (5) are usually referred to as the interaction strength constants which in practice are used to keep term inside the square root positive and set the surface tension, as the latter is tied to $(\mathcal{G}_1+8\mathcal{G}_2)/(\mathcal{G}_1+2\mathcal{G}_2)$~\cite{sbragaglia_generalized_2007}.
In the present work the Peng-Robinson non-ideal equation of state is used for the pressure
\begin{flalign}
    &&
    P=\frac{\rho R T}{1-b\rho} - \frac{a \alpha(T) \rho^2}{1+2b\rho-b^2\rho^2},
    &&
\end{flalign}
 \noindent where
{\small
\begin{flalign}
    &&
    \alpha(T) = {1+(0.37464+1.5422\omega - 0.26992\omega^2)(1-\sqrt{T/T_c})}^2,
    &&
\end{flalign}}

\noindent with $a=0.45724R^2T_{\rm c}^2/P_{\rm c}$, $b=0.0778 R T_{\rm c}/P_{\rm c}$ and $T_{\rm c}$ and $P_{\rm c}$ are the critical temperature and pressure, respectively. The acentric parameter $\omega$ is set to 0.344. In this research, the non-dimensional temperature in the equation of state was set to $T/T_{\rm c}=0.74$, resulting in a density ratio of approximately 220.

\begin{figure}[ht]
	\centering
		\includegraphics[width = 0.9\linewidth]{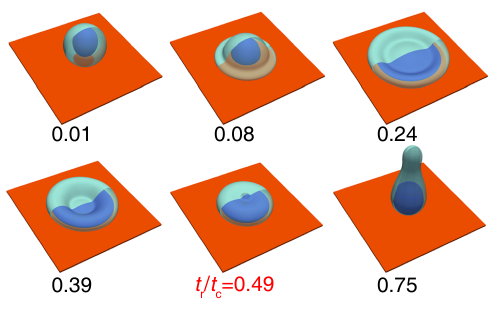}
	\caption{Evolution of lamella over time as obtained from interface tracking simulations at We = 18 and Oh = 0.035. Times are normalized by the total contact time $t_{\rm c}$. The identification process and meaning of the lamella here are discussed in details in Sec.\ \ref{sec:internalflow}.}
	\label{Figure2}
\end{figure}
\subsection{\label{sec:methods:interface}Interface Tracking}
Here, especially for the second part of the study, we will monitor the evolution of different sub-volumes within the droplet over time. For that purpose, the LBM solver for the flow field will be supplemented with an interface tracking algorithm. The interface tracking is performed by solving the conservative Allen-Cahn equation,
\begin{flalign}
    &&
    \partial_t \phi + \bm{\nabla}\cdot \phi \bm{u} - \bm{\nabla}\cdot \left[\mathcal{M}\bm{\nabla}\phi - \bm{n}\frac{4\phi(1-\phi)}{\mathcal{W}}\right] = 0,
    &&
\end{flalign}
where $\phi$ is the order-parameter, $\mathcal{M}$ is the mobility, $\mathcal{W}$ is the interface-thickness and $\bm{n}$ is the unit vector normal to the interface as
\begin{flalign}
    &&
    \bm{n}=\frac{\bm{\nabla}\phi}{\lvert\lvert \bm{\nabla}\phi \lvert\lvert}.
    &&
\end{flalign}
This system results in a conservative advection of the order-parameter space (representing the diffuse interface) with the fluid velocity field. While this equation can be solved using any of the available different space/time discretization strategies, in the context of the present study a modified advection-diffusion LBM scheme was used. Details of the implementation of this solver can be found in previous research \cite{hosseini_lattice_2021}. The interface tracking algorithm is illustrated in Fig.~\ref{Figure2}.
\subsection{\label{sec:methods:tracers}Tracer Particles}
In addition to interface tracking, tracer particles, distributed within the drop will also be used the assess the internal dynamics. For the tracer-particles, they are considered to be mass-less point particles. As such, the evolution equation of a given particle $i$ reduces to
\begin{flalign}
    &&
    \frac{d \bm{r}_i(t)}{dt} = \bm{v}_i(t), 
    &&
\end{flalign}
where $\bm{r}_i$ and $\bm{v}_i$ are particle position and velocity, respectively. Given the lack of inertia and volume, it is readily shown that the Maxey-Riley equation for the particle velocity reduces to $\bm{v}_i(t)=\bm{u}(\bm{r}_i, t)$, where the latter is the fluid velocity at the location of the particle at the considered time \cite{m_kuerten_point-particle_2016}. The particle space/time evolution equation was solved via integration along the path-line and a two-step Runge-Kutta time-marching algorithm. The off-grid fluid velocities needed for the Lagrangian solver were interpolated via third-order Lagrange polynomials from neighboring grid-points.
\section{\label{sec:transitiontime}Weber-independent transition time}
\begin{figure}[!h]
	\centering
		\includegraphics[width = 0.9 \linewidth]{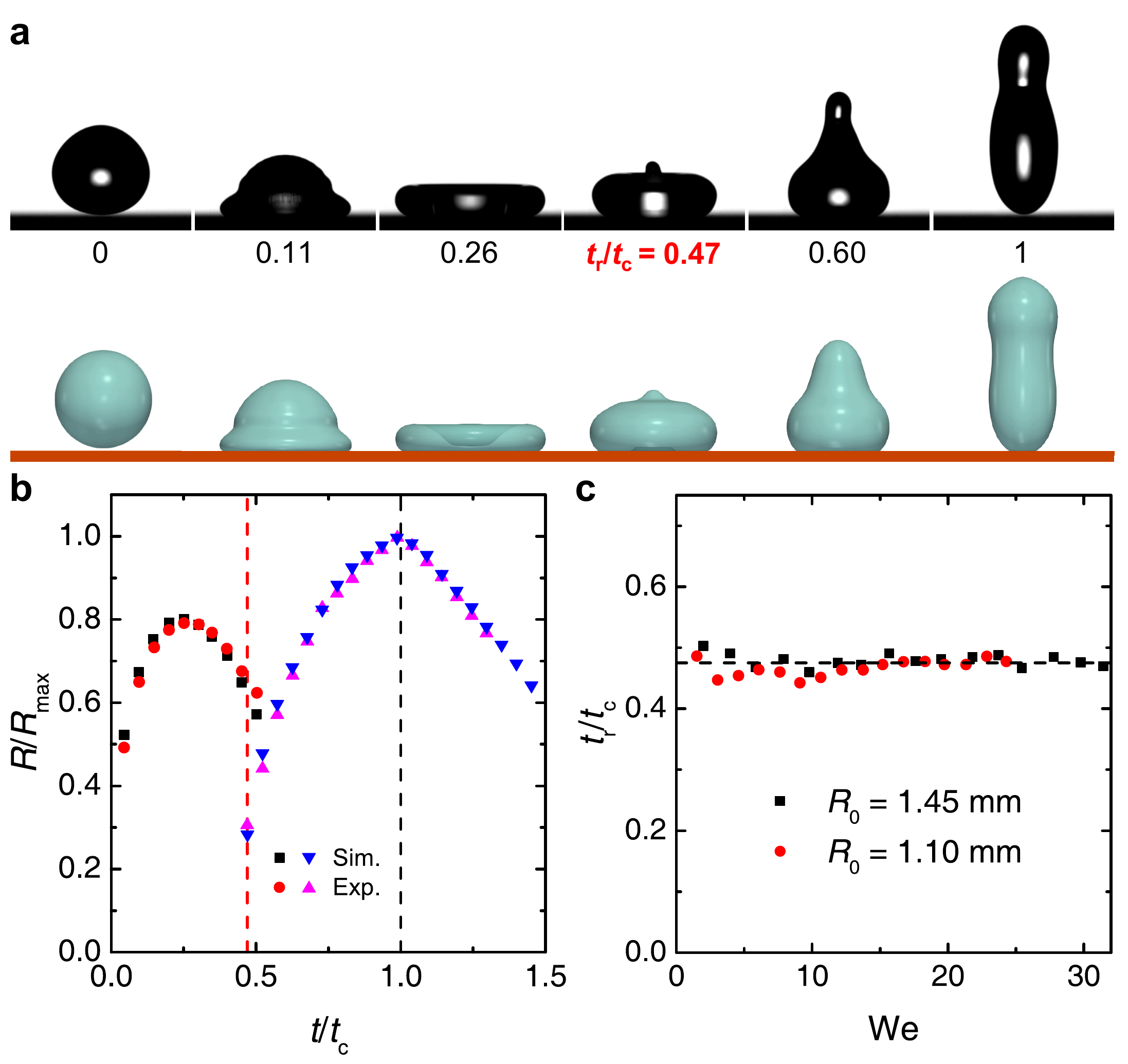}
	\caption{(a) Snapshots showing different stages of drop impact on a SHS at ${\rm We}=5.9$ from both the  experiments and simulations. (b) Evolution of the drop sizes in the radial (red and black) and axial (blue and magenta) directions over time at We = 5.9 from both the  experiments and simulations. (c) Normalized transitions time $\tilde{t}_{\rm r}=t_{\rm r}/t_{\rm c}$ as a function of We at two drop sizes. The dashed black line is the average transition time from the experiments at $\tilde{t}_{\rm r}=0.475$.}
	\label{Figure3}
\end{figure}

\begin{figure}[!ht]
	\centering
	\includegraphics[width = 0.9\linewidth]{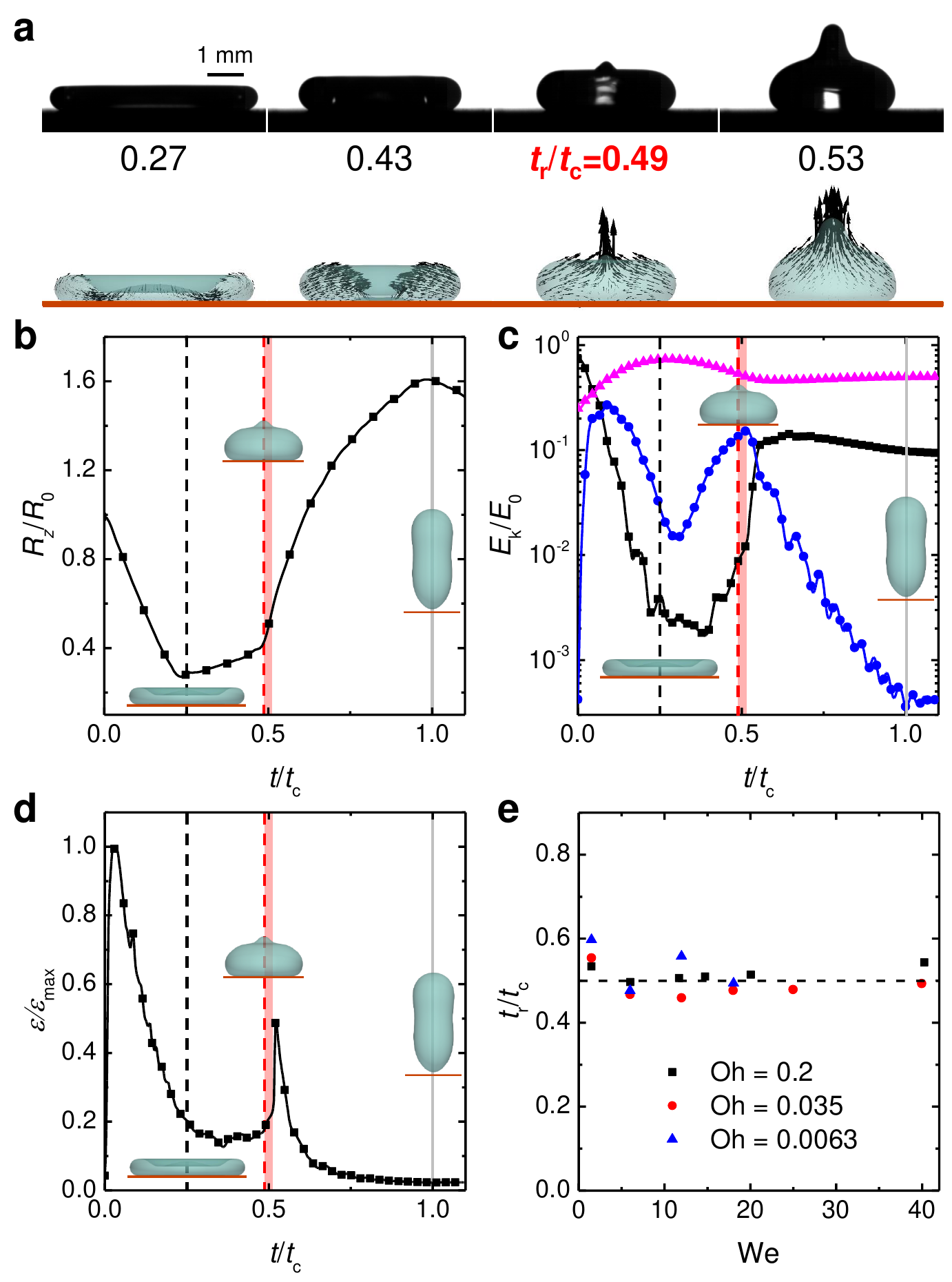}
	\caption{(a) Selected snapshots from simulation (lower panel) showing the time evolution of a drop bouncing and the arrows indicate the velocity field, and the experimental results (upper panel) are presented for comparison. (b) Drop thickness in the axial direction $R_z$ normalized by the initial diameter. (c) Evolution of surface energy (magenta line with triangle markers) and kinetic energy as a function of $t/t_{\rm c}$ in the radial (blue lines with circular markers) and axial (black line with square markers) directions. (d) Viscous dissipation rate as a function of $t/t_{\rm c}$. The black dashed and gray solid lines represent, respectively, the maximum spreading and contact times, while the red dashed line indicates the transition time. The red patch is the transition \emph{interval} obtained from the kinetic energy. Simulation data shown in (a), (b), (c) and (d) are at ${\rm We}=18$ and ${\rm Oh}=0.035$. (e) Transition time $\tilde{t}_{\rm r}$ for different ${\rm Oh}$ and ${\rm Oh}$ as obtained from simulations. The dashed black line is the average transition time at $\tilde{t}_{\rm r}=0.51$.}
    \label{Figure4}
\end{figure}

\begin{figure}[!ht]
	\centering
	\includegraphics[width = 0.6\linewidth]{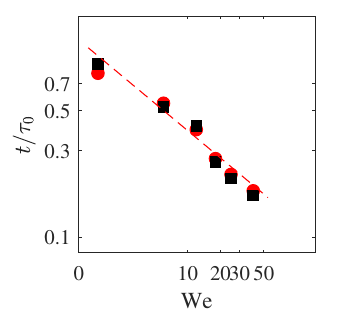}
	\caption{Time of first dissipation rate peaks from simulations. Red circles are from Oh=0035 and black squares from Oh=0.2. The dashed line represents the $-1/2$ slope.}
    \label{Figurefirstdiss}
\end{figure}

We consider drop impact on a flat SHS with an apparent contact angle of $166^{\circ}\pm 3.2^{\circ}$. Defining the inertia-capillarity time as $\tau_0 = \sqrt{\rho R_0^3/\gamma}$, we first verified that the contact time scales as $t_c = (2.5\pm0.1)\tau_0$ for the Weber number range $1.5<{\rm We}<40$, consistent with previous observations \cite{richard_contact_2002,gauthier_water_2015}. Below, all times are reported in units of contact time, $\tilde{t}=t/t_{\rm c}$. Fig.~\ref{Figure3}a visualizes a typical drop impact on a SHS at ${\rm We}=5.9$, where excellent agreement between experiments and simulations is observed.\\
The impact process can be decomposed into multiple stages. This includes a first inertia-dominated spreading stage starting immediately after impact and continuing all the way until spreading is slowed down and eventually stopped at maximum spreading due to surface tension. It is interesting to note, and this will be further discussed in later sections, that at this point the drop consists of a flat thin disk we will refer to as the lamella, surrounded b a donut-like sub-volume we will call the rim. This maximum spreading state is then followed by the retraction phase ending with the appearance of the Worthington jet~\cite{worthington1877}, and axial stretching ending with the contact time. Once the drop reached the maximum spreading at $\tilde{t}_{\rm max}=0.26$, the retraction starts and is characterized by swelling of the peripheral rim. At $t_r/t_c=0.47$, the lamella has collapsed and been entirely absorbed into the receding rim. The collapse of the lamella and subsequent self collision of the rim at $\tilde{t}_{\rm r}=0.47$ leads to the formation of a jet-like structure at the center in the axial direction, also known as the Worthington jet~\cite{doi:10.1098/rsta.1897.0005,doi:10.1098/rsta.1900.0016}. The appearance of this structure, as will be further detailed in following sections, marks the start of a transfer of momentum/energy from radial to axial. In the context of the present study, we term this instance the \emph{transition} time. Finally, the axially elongated drop leaves the surface at $\tilde{t}_{\rm c}=1$.\\
Transition from the predominantly radial to axial motion at $\tilde{t}_{\rm r}$ highlights two stages in Fig.~\ref{Figure3}b, where the lateral extent of the drop was monitored  from the initial contact till $\tilde{t}_{\rm r}$ while the axial elongation was traced thereafter. The drop reaches its maximum radial extent at $\tilde{t}_{\rm max}=0.26$, about half-way through the first stage. Once the drop has retracted, the motion switches to predominantly axial, which is represented by pronounced change in the drop size in the axial direction. Note that, the maximum axial elongation, is reached half-way through the second stage.\\
Fig.~\ref{Figure3}c shows the transition time $\tilde{t}_r$ over a wide range of Weber numbers with two different drop sizes. The results show that this characteristic time, like the contact time, is Weber-independent and also systematically accounts for half of it. It is interesting to note that, though not the focus of the discussion, similar results were also reported in \cite{zhang2022impact}.\\
For a more quantitative understanding, simulations were run with the LBM~\cite{mazloomi_m_entropic_2015,hosseini2022entropic} over $1.5<\hbox{We}<40$ for ${\rm Oh}=0.2$, $0.035$ and $0.0063$, with {${\rm Oh}=\sqrt{\rm We}/{\rm Re}$}, where the Reynolds number is ${\rm Re}=R_0 U_0/\nu$, and $\nu$ is the kinematic viscosity of the liquid. Snapshots of drop impact in Fig. ~\ref{Figure4}a show excellent qualitative agreement between simulations and experiments. By monitoring the half-maximum height of the drop $R_z$ in Fig. ~\ref{Figure4}b, the pronounced change of the slope is clearly seen at the experimentally detected transition time.\\
To better characterize the transition process, we monitor the radial $E_{\rm r}$ and axial $E_{\rm a}$ components of kinetic energy and surface energy in Fig. ~\ref{Figure4}c. At $\tilde{t}=t/t_{\rm c}=0.48$, the radial kinetic energy reaches maximum and starts a sharp descent while the axial component $E_a$ follows the opposite trend. This marks the onset of transition which nears completion at $\tilde{t}=0.53$. Two key observations can be made here: (a) The transition \emph{interval} falls half-way between first and last contact with the SHS and (b) is a relatively fast process as compared to spreading and retraction, which can be considered instantaneous on the overall scale of bouncing. Given that transition is characterized by a self-imploding rim leading to considerable velocity gradients, especially at the center of the drop, one would expect viscous dissipation rate to be affected by the process. The viscous energy dissipation rate~\cite{mazloomi2016simulation} can be computed as
\begin{flalign}
\label{eq:dissipation_rate}
    &&
    \epsilon(t) = \frac{\rho \nu}{2}\int_{V_{\rm drop}} {(\nabla \bm{u} + {\nabla \bm{u}}^{\dagger})}^2 dV,
    &&
\end{flalign}
where $\bm{u}$ is the local fluid velocity. The dissipation rate, as shown in Fig. \ref{Figure4}d, points to two peaks. The first instance comes up right after initial impact and corresponds to the dissipation stemming from the impact impulse. A similar first peak in normal force experience by the substrate is also reported in~\cite{zhang2022impact}. Based on a simple geometrical argument one expects to see a peak in impact force when the largest cross-section of the drop has reached the substrate. This in turn means that this \emph{time} would scale with the inertial time-scale, i.e. $R_0/U_0$, which is confirmed in~\cite{zhang2022impact}. Our studies confirm that this time coincides with the first peak in dissipation rate of the droplet, with a similar scaling, as shown in Fig.~\ref{Figurefirstdiss}.
The second peak coincides with the appearance of the Worthington jet and the directional transfer of kinetic energy. Transition time based on the dissipation rate at the second peak shown with dashed red lines in Fig. \ref{Figure4}d agrees well with the visual identification in Fig. ~\ref{Figure4}e. Thus in the regime of interest, i.e., $1.5<\hbox{We}<40$ and $\hbox{Oh}<1$ where the rim forms and maintains integrity throughout impact, the viscous dissipation rate, similar to the second impulse peak detected in \cite{zhang2022impact} provides an unambiguous measurement of the transition time, that accounts systematically for half of the contact time. \\
Classically, the independence of the contact time on Weber number and the scaling {$t_c\sim \sqrt{\rho R_0^3/\gamma}$} were explained by viewing the overall bouncing as a single harmonic mode~\cite{richard_contact_2002,okumura_water_2003}, with an equivalent spring stiffness $\gamma$ and mass $\rho R_0^3$. While the spring-mass model can also be used to explain the Weber-independent transition time, as the half-way point of a full oscillation period, in the next section we will try to identify characteristic volumes proper to the spreading/retraction phases and discuss another feature observed in our experiments, namely the Weber-independence of the volume partition between two geometrical features of the drop at maximum spreading.
\section{\label{sec:internalflow}Drop internal flow and mass partition}
The analysis of Weber-number-independent contact time~\cite{richard_contact_2002,okumura_water_2003} was extended to more complex dynamics to explain reduced contact times on macro-structured SHSs by identifying characteristic sub-volumes, termed blobs, governing dynamics of the motion. For instance, in \cite{gauthier_water_2015}, at maximum spreading, the drop was represented as connected blobs while the estimate for the contact time followed the mass partition among the blobs. Recently, the blob picture explained yet a different scenario, ring-bouncing on a superhydrophobic defect \cite{chantelot_water_2018}. In all cases, a necessary condition for identification of a characteristic volume/length is its Weber-independence. Since transition time is also Weber-independent, a decomposition that follows a mass partition, should in principle, be applicable. For that purpose and to identify characteristic volumes/scales proper to the spreading/retraction phase we look at the maximum spreading state where two sub-volumes, i.e. rim and lamella, can be clearly distinguished.
\begin{figure}[h]
	\centering
		\includegraphics[width=0.9\linewidth]{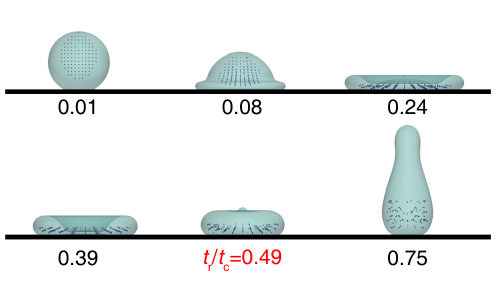}
	\caption{Evolution of point particles location over time as obtained from simulations at We=18 and Oh=0.035. Times are normalized by the total contact time $t_{\rm c}$.}
	\label{Figure5}
\end{figure}
 We first explored the evolution of the sub-volume resulting in the lamella at maximum spreading over time. In order to visualize its history from initial impact up to the transition time, the lamella at the maximum spreading was monitored using an interface tracking and mass-less tracer-particles. First, the impacting drop was uniformly seeded with tracer-particles monitored forward in time. At maximum spreading, all particles within the lamella were identified and traced {\it backward} in time to the initial configuration (Fig. ~\ref{Figure5}). 
An envelope over these  particles was used to identify the sub-volume $V_{\rm lam}$ which would eventually become the lamella forward in time. 
In a second run, the sub-volume was monitored with both tracer-particles and interface tracking, as shown in Fig. ~\ref{Figure6}. Visualization in the figure confirms that the lamella expands and retracts laterally, closely following the radial motion of the drop.
\begin{figure}[h]
	\centering
		\includegraphics[width=0.9\linewidth]{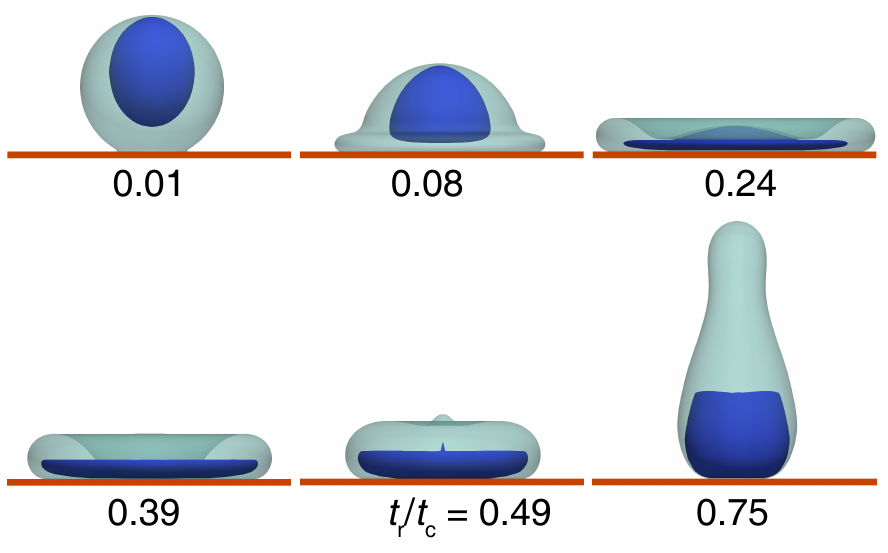}
	\caption{Evolution of the drop shape and of the lamella (dark blue surface) from simulation at ${\rm We}=18$ and ${\rm Oh}=0.035$.}
 \label{Figure6}
\end{figure}
The previous geometric argument was supplemented by a study of flow structure at maximum spreading, e.g. the q-criterion. The q-criterion is defined as the second invariant of the velocity gradient tensor as,
\begin{flalign}
    &&
    q = \frac{1}{2}\left[ {({\rm tr}(\bm{\nabla}\bm{u}))}^2 - {\rm tr}(\bm{\nabla}\bm{u}:\bm{\nabla}\bm{u})\right],
    &&
\end{flalign}
where tr indicates the trace of a matrix and ":" is the Frobenius product of two matrices. Positive values of the $q$-criterion indicate domination of vorticity over strain rate. The $q$-criterion field in Fig. \ref{Figure8}a shows a vortical structure in the rim, in agreement with analogous experiments \cite{clanet_maximal_2004} and simulations \cite{wildeman_visser_sun_lohse_2016}.
\begin{figure}[ht]
	\centering
		\includegraphics[width=0.9\linewidth]{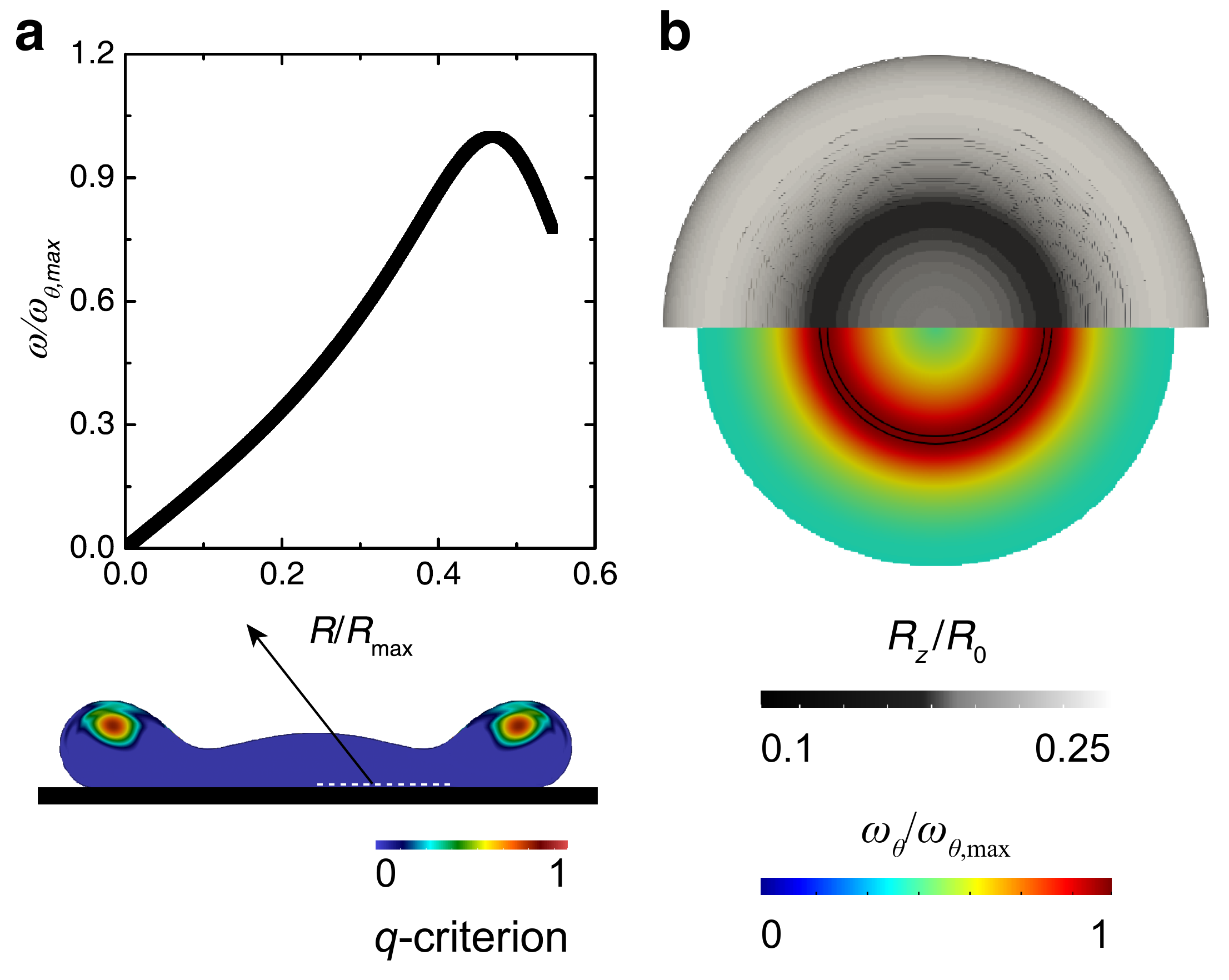}
	\caption{(a) $q$-criterion field in drop at maximum spreading and azimuthal component of vorticity $\omega_\theta$ normalized by its maximum value $\omega_{\theta, max}$, in the boundary layer close to the solid substrate. Inset below the drop shows the velocity distribution in the neck region. (b) Top view of the drop at maximum spreading with height distribution (top half) and azimuthal vorticity distribution (bottom half) at drop interface with substrate. The maximum vorticity region is shown with black contours. Simulation in (a) and (b) is at ${\rm We}=18$ and ${\rm Oh}=0.035$.}
    \label{Figure8}
\end{figure}
The lamella can be assimilated to an expanding boundary layer \cite{schroll_impact_2010,eggers_drop_2010}. Specifically, the initial axial momentum transforms into a radially expanding flow via the no-flux boundary at the substrate surface \cite{schroll_impact_2010}. Due to viscous effects in the drop and the no-slip boundary condition at the wall, one expects vorticity to be generated in the boundary layer, formed close to the substrate. 
To verify the existence of this expanding boundary layer, the distribution of the azimuthal component of the vorticity vector, $\omega_\theta = \partial_r u_z - \partial_z u_r$ with $u_z$ and $u_r$ the axial and radial components of the velocity vector, was computed along the radial axis at the bottom of the lamella, as shown in Fig.  \ref{Figure8}b. Farther from the center, the expansion speed increases, leading to larger gradients and therefore larger vorticity up to the neck region, {i.e.} the onset of the rim, where the flow slows down. A look at the evolution of the volume located in the lamella at maximum spreading and flow structure indicate that the lamella is the characteristic volume associated with the spreading-retraction steps forming the transition time. For the lamella to be the characteristic volume/size, just like the transition time, it must be Weber-independent. Based on apparent differences in flow patterns in the lamella and rim through the spreading process, the volume partition between the rim and lamella was estimated.
\begin{figure}[h]
	\centering
		\includegraphics[width=0.75\linewidth]{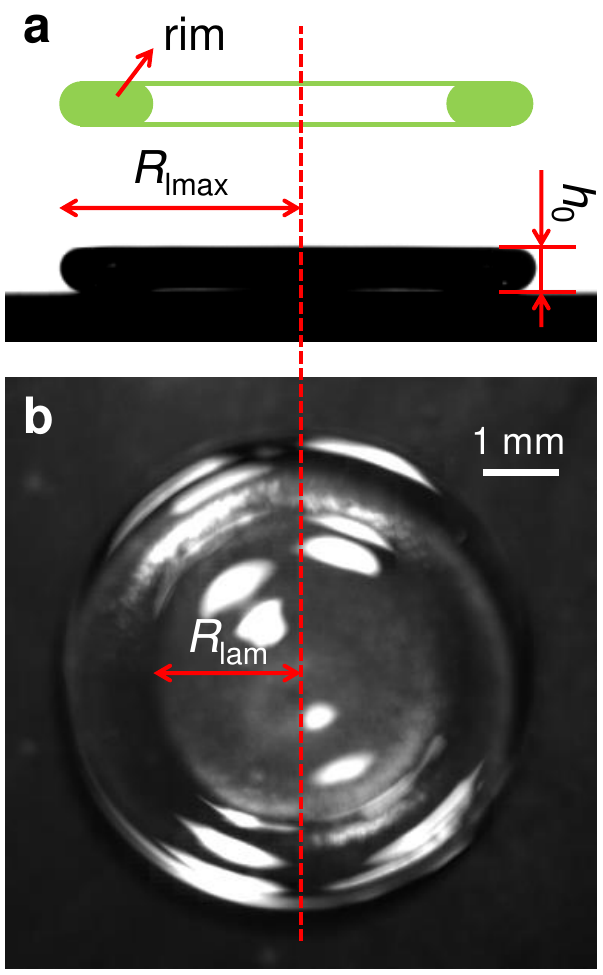}
	\caption{(a) Side- and (b) top-view images at maximum drop spreading at We = 19.7. Here, $R_{\rm lmax}$, $R_{\rm lam}$ and $h_0$ are the maximum spreading radius, lamella radius and thickness of the rim, respectively.}
    \label{Figure7}
\end{figure}
In experiments, the rim was approximated by a torus based on the top-view of the drop at maximum extension (Fig. ~\ref{Figure7}), while in simulations, the interface between was identified via the neck at the outer circumference of the lamella, as shown in Figs. ~\ref{Figure8}a and ~\ref{Figure8}b. Fig. ~\ref{Figure9} shows that the relative volume of the rim constitutes $V_{\rm rim}/V_0= 0.75\pm 0.05$, i.e. the lamella account for a quarter of the drop total volume, and is Weber-independent. This indicates that the prediction by the spring-mass argument holds, $\sqrt{V_{\rm blob}/V_0} \approx {t_{\rm r}/t_{\rm c}} \approx 1/2$, provided the equivalent volume of the blob associated with transition time is that of the lamella at maximum extension, $V_{\rm blob}/V_0 = V_{\rm lam}/V_0 \approx 1/4$.
\begin{figure}[h]
	\centering
		\includegraphics[width=0.9\linewidth]{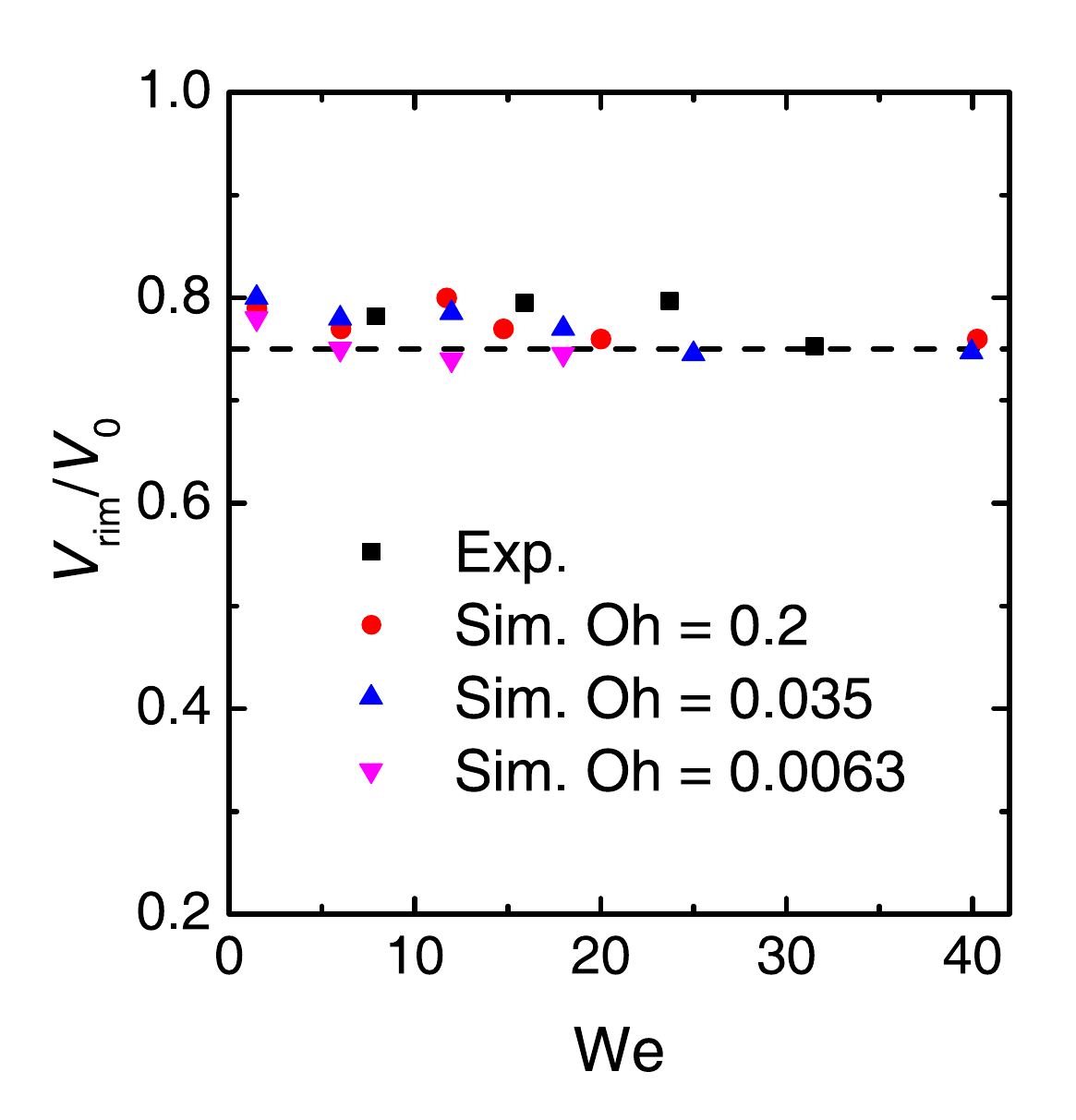}
	\caption{Volume partition in the drop obtained from both the experiments and simulations. The dashed black line is the average ratio between the volume of the rim and drop volume from both numerical and experimental data at $V_{\rm rim}/V_0=0.77$.}
    \label{Figure9}
\end{figure}
This observation, i.e. Weber-invariance of the volume partition between rim and lamella at maximum spreading, along with specific flow dynamics in each sob-volumes point to the fact that the lamella can be associated as the characteristic volume to the spreading/retraction phase. This is further confirmed by the fact the relative volume of the lamella at maximum spreading perfectly agrees with previously measured transition times throught spring-mass model.
\section{\label{sec:conclusion}Conclusion}
In summary the transition time, visually identifiable via the appearance of the Worthington jet, is shown to be a Weber-independent characteristic time of the drop impact process. This characteristic time was observed to mark the transition of droplet dynamics from radial to axial. The transition time is robust, that is, independent of impact kinetic energy (or Weber number) and accounts systematically for half the contact time. Detailed studies of flow dynamics showed that this instance can be identified through a peak in dissipation rate within the droplet. A first peak in dissipation rate, corresponding to a peak in droplet impact on the substrate was also identified and shown to be in excellent agreement with observations reported in \cite{zhang2022impact}. In addition a detailed study of the flow structure and drop shape at maximum spreading state showed an additional Weber-independent variable, that is the volume partition at maximum spreading between the rim and lamella. Both experiments and simulations confirmed that at maximum spreading the lamella account for a quarter of the drop total volume. These observations allowed us to identify the lamella as the blob associated to the spreading/retraction process and explain the fact that the latter systematically account for half the contact time through the mass-spring analogy.
\section{\label{sec:acknowledgements}Acknowledgements}
This work was supported by the National Key R\&D Program of China (No. 2022YFB4602401), the National Natural Science Foundation of China Grants (No. 52075071), the European Research Council (ERC) grant (No. 834763-PonD), the Swiss National Science Foundation (SNSF) grants (No. 200021-172640 an 200021-228065) and the Hong Kong Research Grants Council Grant (No. 11219219). Computational resources at the Swiss National Super Computing Center (CSCS) were provided under grants No. s897, s1066 and s1286.
\bibliography{references}

\end{document}